\documentclass[preprint,showpacs,showkeys,preprintnumbers,amsmath,amssymb]{revtex4}
\usepackage{amssymb}
\usepackage[dvips]{graphicx}

\begin{document}

\title{New symmetry reductions related with the residual symmetry of Boussinesq equation}

\author{ Xi-zhong Liu, Jun Yu, Bo Ren }

\affiliation{Institute of Nonlinear Science, Shaoxing University, Shaoxing 312000, China}
\begin{abstract}
The B\"{a}klund transformation related symmetry is nonlocal, which is hardly to apply in constructing solutions for nonlinear equations. In this paper, we first localize nonlocal residual symmetry to Lie point symmetry by introducing multiple new variables and obtain new B\"{a}klund transformation. Then, by solving out the general form of localized the residual symmetry, we reduce the enlarged system by classical symmetry approach and obtain the corresponding reduction solutions as well as related reduction equations. The localization procedure provides a new way to investigate interaction solutions between different waves.
\end{abstract}



\maketitle
\section{Introduction}
It is known that the the physical laws ruling our world is essentially nonlinear, which explains why almost all of the fundamental equations of natural science, including physics, chemistry, biology, astrophysics, etc., are nonlinear. To discover these laws and write down the corresponding equations is important, while to study these nonlinear equations by varied methods through which to reveal the underlying meaning is equally important. Nonlinear equations are usually hard to solve out explicitly, so perturbation, asymptotic and numerical methods are often used, with much success, to obtain approximate solutions of these equations. However, there is also much current interest in obtaining exact analytical solutions of nonlinear equations. Among which, the study on integrable system plays an important role and various effective methods have been developed to study its different properties.

Since the Lie group theory was introduced by Sophus Lie to study differential equations \cite{lie}, the study of Lie group has always been an important subject in mathematics and physics. If the nonlinear equations has Lie point symmetry group, one can lower dimensions of partial differential equations (PDEs) by using both classical and non-classical Lie group approaches \cite{olver,bluman}. This provides a method for obtaining exact and special solutions of a given equation in terms of solutions of lower dimensional equations, in particular, ordinary differential equations. B\"{a}klund transformation (BT) is another powerful method to search for new solutions of PDEs through known ones. But the interaction between different waves is hard to study by the original BT because it is very difficult to solve when the seed solutions is taken as nonconstant nonlinear waves such as the cnoidal waves and Painlev¡äe waves. In this paper, we no longer concentrate on the traditional BT itself but switch to study its infinitesimal transformation, i.e. its symmetry and use this symmetry to construct new exact solutions. However, the BT related symmetry is nonlocal \cite{97lou}, which is hardly used for reducing the corresponding equation. One of the best way to concur this difficulty is to localize it by introducing new dependent variables, such that the symmetry become localized in the new enlarged system \cite{xiaonan,huxiao}.

In this paper, we consider the well-known Boussinesq equation in the form
\begin{equation}\label{bouss}
u_{tt}+u_{xxxx}+6u_x^2+6uu_{xx}=0.
\end{equation}
The Boussinesq equation arises in several physical applications, such as propagation of long waves in shallow water \cite{bouss1,bouss2,whit}, one-dimensional nonlinear lattice-waves \cite{toda,zab}, vibrations in a nonlinear string \cite{zakh} and ion sound waves in a plasma \cite{infeld,scott}, etc. For the Boussinesq equation, a considerable number of explicit solutions, especially the multisoliton solutions and periodic solutions, have been obtained by various methods including inverse scattering transformation, the symmetry method, the Hirota bilinear method, the Darboux transformation (DT), the B\"{a}klund transformation, the tanh expansion method, etc. 

The paper is organized as follows. In section 2, we obtain the residual symmetry by using the Painlev\'{e} truncated expansion method then we localize it by introducing multiple new variables.
By the way, the finite group transformation theorem is consequently obtained by Lie's first principle. In section 3, the general form of Lie point symmetry group for the enlarged Boussinesq system is obtained and the similarity reductions for the prolonged system is considered according to the standard Lie point symmetry approach with some special explicit solutions are given. The last section is devoted to a short summary and discussion.

\section{localization of residual symmetry and the related B\"{a}cklund transformation }

Since the Boussinesq equation has the Painlev\'{e} property, i.e. it is Painlev\'{e} integrable, there exist a truncated transformation
\begin{equation}\label{genpain}
u=\sum_{i=0}^{\alpha}{u_i\phi^{i-\alpha}},
\end{equation}
for positive integer $\alpha$. By balancing the nonlinear term and dispersion term in Eq. \eqref{bouss}, we have $\alpha=2$. Hence the truncated Painlev\'{e} expansion reads
\begin{equation}\label{pain1}
u=\frac{u_0}{\phi^2}+\frac{u_1}{\phi}+u_2.
\end{equation}
It an be easily verified that the residue $u_1$ of the truncated Painlev\'{e} expansion \eqref{pain1} with respect to the singular manifold $\phi$ is a symmetry of \eqref{bouss} with the solution $u_2$. To prove this conclusion, we substitute \eqref{pain1} into the Eq. \eqref{bouss} and get
\begin{multline}\label{subspain1}
6u_{2x}^2+6u_2u_{2xx}+u_{2xxxx}+u_{2tt}+\phi^{-1}(6u_1u_{2xx}+12u_{1x}u_{2x}+u_{1xxxx}+u_{1tt}
+6u_2u_{1xx})\\+\phi^{-2}(-4u_{1x}\phi_{xxx}+6u_0u_{2xx}-12u_2u_{1x}\phi_x-12u_1\phi_xu_{2x}-4u_{1xxx}
\phi_x-6u_{1xx}\phi_{xx}+6u_1u_{1xx}\\+u_{0xxxx}+6u_{1x}^2+u_{0tt}-u_1\phi_{xxxx}+12u_{0x}u_{2x}
-2u_{1t}\phi_t-u_1\phi_{tt}-6u_2u_1\phi_{xx}+6u_2u_{0xx})\\+\phi^{-3}(2u_1\phi_t^2+12u_{0x}u_{1x}+6u_0u_{1xx}
-2u_0\phi_{xxxx}+12u_2u_1\phi_x^2-24u_0\phi_xu_{2x}-12u_2u_0\phi_{xx}\\-4u_{0t}\phi_t-12u_{0xx}
\phi_{xx}+6u_1u_{0xx}-8u_{0x}\phi_{xxx}+6u_1\phi_{xx}^2-8u_{0xxx}\phi_x-24u_1u_{1x}\phi_x
-24u_2u_{0x}\phi_x\\+8u_1\phi_x\phi_{xxx}-6u_1^2\phi_{xx}+12u_{1xx}\phi_{x}^2+24u_{1x}\phi_x
\phi_{xx}-2u_0\phi_{tt})+\phi^{-4}(6u_0u_{0xx}-36u_0u_{1x}\phi_x\\+6u_0\phi_t^2-18u_0u_1\phi_{xx}-36u_1\phi_x^2
\phi_{xx}+72u_{0x}\phi_x\phi_{xx}+6u_{0x}^2+36u_2u_0\phi_x^2-36u_1u_{0x}\phi_x+18u_0\phi_{xx}
^2\\+18u_1^2\phi_x^2+36u_{0xx}\phi_x^2-24u_{1x}\phi_x^3+24u_0\phi_x\phi_{xxx})
+\phi^{-5}(-12u_0^2\phi_{xx}-96u_{0x}\phi_x^3+72u_0u_1\phi_x^2+24u_1\phi_x^4\\-48u_0u_{0x}\phi_x-144u_0\phi_x^2
\phi_{xx})+60\phi^{-6}u_0\phi_x^2(2\phi_x^2+u_0).
\end{multline}
Vanishing the coefficients of $\phi^0$ and $\phi^{-1}$ in \eqref{subspain1}, we see that $u_1$ satisfies the linearized Boussinesq equation with $u_2$ being the solution, thus our conclusion is proved.

In order to determine the residual symmetry $\sigma^u=u_1$ and the Schwarz form of Boussinesq equation, we proceed to vanish the coefficients of $\phi^{-6}$, $\phi^{-5}$ and $\phi^{-4}$ in \eqref{subspain1} then we have
\begin{equation}\label{u0v0}
u_0 = -2\phi_x^2,\quad u_1 = 2\phi_{xx}
\end{equation}
and
\begin{equation}\label{u2}
u_2 =-\frac{1}{6}\phi_x^{-2}(-3\phi_{xx}^2+4\phi_x\phi_{xxx}+\phi_t^2).
\end{equation}
Now by substituting $u_0$, $u_1$ and $u_2$ in Eqs. \eqref{u0v0} and \eqref{u2} into Eq. \eqref{subspain1} meanwhile vanishing the coefficient of $\phi^{-3}$, we get the schwarz form of Boussinesq equation
\begin{equation}\label{schwarzform}
\bigg[\{\phi;x\}+\frac{1}{2}\frac{\phi_t^2}{\phi_x^2}\bigg]_x+\bigg(\frac{\phi_t}{\phi_x}\bigg)_t=0,
\end{equation}
with $\{\phi;x\}=\frac{\phi_{xxx}}{\phi_x}-\frac{3}{2}\frac{\phi_{xx}^2}{\phi_x^2}$.
Eq. \eqref{schwarzform} is form invariant under the M\"{o}bious transformation
\begin{equation}
\phi\to \frac{a_1\phi+b_1}{a_2\phi+b_2},\, a_1a_2 \neq b_1b_2,
\end{equation}
which means Eq. \eqref{schwarzform} possess three symmetries $\sigma_{\phi}=d_1$,\ $\sigma_{\phi}=d_2\phi$ and
\begin{equation}\label{symmob}
\sigma_{\phi}=d_3\phi^2
\end{equation}
with arbitrary constants $d_1,\,d_2$ and $d_3.$ Apparently, the residual symmetry $\sigma^u=u_1$ is related with the M\"{o}bious transformation symmetry \eqref{symmob} by the linearized equation of \eqref{u2}.

It is clear that the residual symmetry 
\begin{equation}\label{residualsym}
\sigma_u=2\phi_{xx}
\end{equation}
with $\phi$ satisfying Eq. \eqref{schwarzform} is nonlocal and this symmetry is just the generator of the  B\"{a}cklund transformation \eqref{pain1}. In order to study the transformation properties of Boussinesq equation by this symmetry, we have to  localize it to Lie point symmetry by introducing the following new variables
\begin{equation}\label{gh1}
g\equiv \phi_x,
\end{equation}
\begin{equation}\label{gh2}
 h\equiv g_x,
\end{equation}
\begin{equation}\label{mphi}
m\equiv \phi_t.
\end{equation}
All the symmetries corresponding to different variables are related each other by the linearized equations of \eqref{bouss}, \eqref{schwarzform},\eqref{gh1}, \eqref{gh2} and \eqref{mphi}, i.e.,
\begin{subequations}\label{linear}
\begin{equation}
\sigma_{\phi,x}-\sigma_g=0,
\end{equation}
\begin{equation}
\sigma_{g,x}-\sigma_h=0
 \end{equation}
\begin{equation}
 \sigma_{\phi,t}-\sigma_m=0,
\end{equation}
\begin{equation}
 -3\phi_x\phi_{xx}\sigma_{\phi, xx}-2\phi_x\sigma_{\phi, x}\phi_{xxx}+2\phi_x^2\sigma_{\phi, xxx}+\phi_x\phi_t\sigma_{\phi, t}+3\sigma_{\phi, x}\phi_{xx}^2-\sigma_{\phi, x}\phi_t^2+3\sigma_u\phi_x^3=0
\end{equation}
\begin{equation}
 \sigma_{u,tt}+ \sigma_{u,xxxx}+12u_x \sigma_{u,x}+6\sigma_uu_{xx}+6u \sigma_{u,xx}=0,
\end{equation}
\begin{equation}\nonumber
-\sigma_{\phi,xx}\phi_t^2-2\sigma_{\phi,t}\phi_{xx}\phi_t+2\phi_x\phi_{xxxx}
\sigma_{\phi,x}+\phi_x^2\sigma_{\phi,xxxx}+9\phi_{xx}^2\sigma_{\phi,xx}
-4\sigma_{\phi,x}\phi_{xx}\phi_{xxx}
\end{equation}
\begin{equation}-4\phi_x\sigma_{\phi,xx}\phi_{xxx}-4\phi_x\phi_
{xx}\sigma_{\phi,xxx}+2\phi_x\phi_{tt}\sigma_{\phi,x}+\phi_x^2\sigma_{\phi,tt}=0
\end{equation}
\end{subequations}
It can be easily verified that the solutions of \eqref{linear} has the form
\begin{subequations}\label{pointsy1}
\begin{equation}
\sigma_{u} =-h,
\end{equation}
\begin{equation}
\sigma_{g} = \phi g,
\end{equation}
\begin{equation}
\sigma_h = g^2+\phi h,
\end{equation}
\begin{equation}
\sigma_{\phi} = \frac{\phi^2}{2},
\end{equation}
\begin{equation}
\sigma_{m} =\phi m,
\end{equation}
\end{subequations}
if $d_3=\frac{1}{2}$ and $d_1=d_2=0$ is fixed for $\sigma_{\phi}$.

The result \eqref{pointsy1} indicates that the residual symmetries \eqref{residualsym} is localized in
the properly prolonged system \eqref{bouss}, \eqref{schwarzform}, \eqref{gh1}, \eqref{gh2} and \eqref{mphi} with the Lie point symmetry vector
\begin{equation}\label{pointV}
V=-h\partial_{u}+g\phi\partial_{g}+(g^2+h\phi)\partial_{h}+m\phi\partial_m
+\frac{\phi^2}{2}\partial_{\phi}.
\end{equation}
In other words, symmetries related to the truncated Painlev\'{e} expansion is just a special Lie point symmetry of the prolonged system.

Now, let us study the finite transformation group corresponding to the Lie point symmetry \eqref{pointV}, which can be stated in the following theorem.

\noindent\emph{ \textbf{Theorem 1.}}
If $\{u,g,h,m,\phi\}$ is a solution of the prolonged system \eqref{bouss}, \eqref{schwarzform}, \eqref{gh1}, \eqref{gh2} and \eqref{mphi}, then so is $\{\hat{u},\hat{g},\hat{h},\hat{m},\hat{\phi}\}$ with
\begin{subequations}\label{pointsy}
\begin{equation}
\hat{u}= u+\frac{2h\epsilon}{\epsilon\phi-2}-\frac{2\epsilon^2g^2}{(\epsilon\phi-2)^2},
\end{equation}
\begin{equation}
\hat{g} = \frac{4g}{(\epsilon\phi-2)^2},
\end{equation}
\begin{equation}
\hat{h} = \frac{4h}{(\epsilon\phi-2)^2}-\frac{8\epsilon g^2}{(\epsilon\phi-2)^3},
\end{equation}
\begin{equation}
\hat{m} = \frac{4m}{(\epsilon\phi-2)^2},
\end{equation}
\begin{equation}
\hat{\phi} =-\frac{2\phi}{\epsilon\phi-2},
\end{equation}
\end{subequations}
with arbitrary group parameter $\epsilon$.

\emph{Proof.} Using Lie's first theorem on vector \eqref{pointV} with the corresponding
 initial condition as follows
\begin{eqnarray}
\\ \frac{d\hat{u}(\epsilon)}{d\epsilon}&=& -\hat{h}(\epsilon),\,\quad \hat{u}(0)=u,\\
\frac{d\hat{g}(\epsilon)}{d\epsilon}&=& \hat{\phi}(\epsilon)\hat{g}(\epsilon),\,\quad \hat{g}(0)=g,\\
\frac{d\hat{h}(\epsilon)}{d\epsilon}&=&\hat{g}(\epsilon)^2+\hat{\phi}(\epsilon)\hat{h}(\epsilon),\,\quad \hat{h}(0)=h,\\
\frac{d\hat{\phi}(\epsilon)}{d\epsilon}&=&\frac{\hat{\phi}(\epsilon)^2}{2},\,\quad \hat{\phi}(0)=\phi,\\
\frac{d\hat{m}(\epsilon)}{d\epsilon}&=&\hat{\phi}(\epsilon)\hat{m}(\epsilon),\,\quad \hat{m}(0)=m,
\end{eqnarray}
one can easily obtain the solutions of the above equations stated in Theorem 1, thus the theorem is
proved.
\section{new symmetry reductions of Boussinesq equation}
Because the nonlocal symmetries cannot be used directly to constructed exact solutions of differential equations, we seek the Lie point symmetry of the prolonged Boussinesq system in the general form
 \begin{equation}\label{vectorv1}
V=X\frac{\partial}{\partial x}+T\frac{\partial}{\partial
t}+U\frac{\partial}{\partial u}+G\frac{\partial}{\partial g}+H\frac{\partial}{\partial h}+
M\frac{\partial}{\partial m}+\Phi\frac{\partial}{\partial \phi},
\end{equation}
which means that the prolonged system \eqref{bouss}, \eqref{schwarzform}, \eqref{gh1}, \eqref{gh2} and \eqref{mphi} is invariant under the following transformation
\begin{equation}
\{x,t,u,g,h,m,\phi\} \rightarrow \{x+\epsilon X,t+\epsilon T,u+\epsilon U,g+\epsilon G,h+\epsilon H,m+\epsilon M,\phi+\epsilon \Phi\}
\end{equation}
with the infinitesimal parameter $\epsilon$.
Equivalently, the symmetry in the form \eqref{vectorv1} can be written as a function form as
\begin{subequations}\label{sigmasy}
\begin{equation}
\sigma_{u} = Xu_{x}+Tu_{t}-U,
\end{equation}
\begin{equation}
\sigma_{g} = Xg_{x}+Tg_{t}-G,
\end{equation}
\begin{equation}
\sigma_h = Xh_{x}+Th_{t}-H,
\end{equation}
\begin{equation}
\sigma_m = Xm_{x}+Tm_{t}-M,
\end{equation}
\begin{equation}
\sigma_{\phi} = X\phi_{x}+T\phi_{t}-\Phi.
\end{equation}
\end{subequations}
Substituting Eq. \eqref{sigmasy} into Eq. \eqref{linear} and eliminating $u_{t}, g_{x}, g_t, h_x, h_t, m_t,\phi_x$ and $\phi_t$ in terms of the prolonged system, we get more than 200 determining equations for the functions $X, T, U, G, H, M$ and $\Phi$. Calculated by computer algebra, we finally get the desired result
\begin{eqnarray}\label{sol}
\nonumber&&X= \frac{c_1}{2}x+c_3, T= c_1t+c_2,U = -c_1u+c_4h,\\\nonumber&&G = -c_4g\phi+\frac{1}{2}
(2c_5-c_1)g, H= -c_4h\phi+(c_5-c_1)h-c_4g^2,\\&& M = -c_4m\phi+(c_5-c_1)m,\Phi = -\frac{c_4}{2}\phi^2+c_5\phi+c_6.
\end{eqnarray}
with arbitrary constants $c_1, c_2, c_3, c_4, c_5, c_6.$ When $c_1=c_2=c_3=c_5=c_6=0$, the obtained symmetry degenerated into the special form in Eq. \eqref{pointV} which includes residual symmetry of Boussinesq equation. By setting $c_2=c_3=c_4=c_5=c_6=0$, we get the scaling transformation symmetry as
\begin{multline}
\sigma_u=\frac{1}{2}c_1xu_x+c_1tu_t+c_1u,\,\sigma_g=\frac{1}{2}c_1xg_x+c_1tg_t+\frac{c_1}{2}g,
\,\sigma_h=\frac{1}{2}c_1xh_x+c_1th_t+c_1h,\,
\\\sigma_m=\frac{1}{2}c_1xm_x+c_1tm_t+c_1m,\,\sigma_{\phi}=\frac{1}{2}c_1x\phi_x+c_1t\phi_t+c_1\phi.
\end{multline}
Other well-known symmetries, like $x$ and $t$ translation invariance symmetries, are also included in the symmetry group.

Consequently, the symmetries in \eqref{sigmasy} can be written as
\begin{eqnarray}\label{sigmauvf}
\nonumber\sigma_{u}&=&\frac{c_1}{2}xu_x+(c_1t+c_2)u_t+c_1u-c_4h,\\
\nonumber\sigma_{g}&=&\frac{c_1}{2}xg_x+(c_1t+c_2)g_t+c_4\phi g+\frac{1}{2}(c_1-2c_5)g,\\
\sigma_h&=&\frac{c_1}{2}xh_x+(c_1t+c_2)h_t+c_4h\phi-(c_5-c_1)h+c_4g^2,\\
\nonumber\sigma_m&=&\frac{c_1}{2}xm_x+(c_1t+c_2)m_t+c_4m\phi-(c_5-c_1)m,\\
\nonumber\sigma_{\phi}&=&\frac{c_1}{2}x\phi_x+(c_1t+c_2)\phi_t+\frac{c_4}{2}\phi^2-c_5\phi
\end{eqnarray}
To give the group invariant solutions of the enlarged system, we have to solve the equations \eqref{sigmauvf} under symmetry constraints $\sigma_u=\sigma_g=\sigma_h=\sigma_m=\sigma_{\phi}=0$, which is equivalent to solve the corresponding characteristic equation
\begin{multline}\label{chac}
\frac{dx}{\frac{c_1}{2}x+c_3}=\frac{dt}{ c_1t+c_2}=\frac{du}{-c_1u+c_4h}=\frac{dg}{-c_4g\phi+\frac{1}{2}
(2c_5-c_1)g}=\frac{dh}{-c_4h\phi+(c_5-c_1)h-c_4g^2}\\=\frac{dm}{-c_4m\phi+(c_5-c_1)m}
=\frac{d\phi}{-\frac{c_4}{2}\phi^2+c_5\phi+c_6}.
\end{multline}

Without loss of generality, we consider the symmetry reductions categorized as the following three
subcases.

\noindent \textbf{Case 1}. $c_i\neq0$ (i=1,\,2,\,4,\,5) and $c_3=c_6=0$.

In this case, by solving Eq. \eqref{chac}, we get
\begin{eqnarray}\label{redusol01}
\phi&=&\frac{2c_5}{2c_5\Delta^{-\frac{2c_5}{c_1}}\Phi+c_4},\,\Delta=\frac{\sqrt{c_1\xi(c_1t+c_2)}}{c_1\xi},\\
\label{redusol02}g&=&\frac{\Delta^{-(\frac{2c_5}{c_1}+1)}G}{\bigg[2c_5\Delta^{-\frac{2c_5}{c_1}}\Phi+c_4\bigg]^2},\\
\label{redusol03}h&=&\frac{4c_5^2\Delta^{(\frac{2c_5}{c_1}-2)}\Phi^2H}{\bigg[c_4\Delta^{\frac{2c_5}{c_1}}+2c_5\Phi\bigg]^2}\nonumber\\&&-
\frac{c_4\Delta^{(\frac{4c_5}{c_1}-2)}G^2}{2c_5^2\bigg[c_4\Delta^{\frac{2c_5}{c_1}}+2c_5\Phi\bigg]^3\Phi}
,\\
\label{redusol04}m&=&\frac{\Delta^{(-\frac{2c_5}{c_1}-2)}M}{\bigg[2c_5\Delta^{-\frac{2c_5}{c_1}}\Phi+c_4\bigg]^2},\\
\label{redusol05}u&=&\frac{ U}{\Delta^2}+\frac{2c_4\Delta^{(\frac{2c_5}{c_1}-2)}\Phi H}{\bigg[c_4\Delta^{\frac{2c_5}{c_1}}+2c_5\Phi\bigg]}\nonumber\\&&-\frac{c_4^2
\Delta^{(\frac{4c_5}{c_1}-2)}G^2}
{8c_5^4\bigg[c_4\Delta^{\frac{2c_5}{c_1}}+2c_5\Phi\bigg]^2\Phi^2},
\end{eqnarray}
where the group invariant variable is $\xi=\frac{c_1t+c_2}{c_1x^2}$, while $U,\, G,\, H,\, M$ and $\Phi$ are all group invariant functions of $\xi$.

To get the symmetry reduction equations for the group invariant functions $U,\, G,\, H,\, M$ and $\Phi$, we first substitute Eqs. \eqref{redusol01}, \eqref{redusol02}, \eqref{redusol03} and \eqref{redusol04} into Eqs. \eqref{gh1},  \eqref{gh2} and \eqref{mphi}, then we have
\begin{equation}\label{redmphi}
4c_5^2\Phi_{\xi}+M=0,
\end{equation}
\begin{equation}\label{redgphi}
8c_5^3\Phi+8c_5^2c_1\xi\Phi_{\xi}-c_1G=0
\end{equation}
and
\begin{equation}\label{redhphi}
-8c_1^2\Phi_{\xi}^2\xi^2+2c_1(3c_1-4c_5)\xi\Phi\Phi_{\xi}+4c_1^2\xi^2\Phi\Phi_{\xi\xi}
+(2c_1c_5-4c_5^2+c_1^2H\Phi)\Phi^2=0.
\end{equation}
Next, we substitute Eqs. \eqref{redusol01} and \eqref{redusol05} into Eqs. \eqref{schwarzform} and \eqref{u2} with $u_2$ substituted by $u$, replacing $M,\, G,\, H$ by Eqs. \eqref{redmphi}, \eqref{redgphi} and \eqref{redhphi} at the same time, we finally get the symmetry reduction equations for $U$ and $\Phi$ as
\begin{multline}\label{reduphi}
192c_1^4\xi^4\Phi_{\xi}^4-96c_1^3(3c_1-4c_5)\xi^3\Phi\Phi_{\xi}^3
+\big[-192c_1^4\xi^4\Phi\Phi_{\xi\xi}+12c_1^2(16c_5^2+7c_1^2+2c_1^2U\\-48c_1c_5)\xi^2\Phi^2+c_1^4\Phi^2\big]
\Phi_{\xi}^2+\big[48c_1^3(-8c_5+3c_1)\xi^3\Phi^2\Phi_{\xi\xi}+64c_1^4\xi^4\Phi^2\Phi_{\xi\xi\xi}
+8c_1c_5(19c_1^2+8c_5^2\\-30c_1c_5+6c_1^2U)\xi\Phi^3\big]\Phi_{\xi}-48c_1^4\xi^4\Phi^2\Phi_{\xi\xi}^2
+48c_1^2c_5(-2c_5+5c_1)\xi^2\Phi^3\Phi_{\xi\xi}+64c_1^3c_5\xi^3\Phi^3\Phi_{\xi\xi\xi}
\\+4c_5^2(4c_5^2-12c_1c_5+5c_1^2+6c_1^2U)\Phi^4=0
\end{multline}
and
\begin{multline}\label{redphi}
\big[4c_1^2(4c_5+c_1)(-8c_5^2-4c_1c_5+3c_1^2)\xi^3+c_1^4(3c_1+4c_5)\xi\big]\Phi_{\xi}^3
+\big[96c_1^3c_5(-2c_5+c_1)\xi^4\Phi_{\xi\xi}\\-32c_1^4(3c_1-4c_5)\xi^5\Phi_{\xi\xi\xi}
-32c_1^5\xi^6\Phi_{\xi\xi\xi\xi}+4c_1c_5(9c_1^3-8c_5^3+72c_5^2c_1+80c_1^2c_5)\xi^2\Phi+c_1^3c_5
(c_1+2c_5)\Phi\big]\Phi_{\xi}^2\\+\big[48(3c_1-4c_5)c_1^4\xi^5\Phi_{\xi\xi}^2+(128c_1^5\xi^6\Phi_{\xi\xi\xi}
-32c_5c_1^2(c_1^2-21c_1c_5-4c_5^2)\xi^3\Phi-4c_1^4c_5\xi\Phi)\Phi_{\xi\xi}
\\-64c_1^3c_5(-2c_5+5c_1)\xi^4\Phi\Phi_{\xi\xi\xi}-64c_1^4c_5\xi^5\Phi\Phi_{\xi\xi\xi\xi}
-4c_5^2c_1(19c_1^2-28c_5^2)\xi\Phi^2\big]\Phi_{\xi}-96c_1^5\xi^6\Phi_{\xi\xi}^3
\\+48c_1^3c_5(2c_5+9c_1)\xi^4\Phi\Phi_{\xi\xi}^2+\big[128c_1^4c_5\xi^5\Phi\Phi_{\xi\xi\xi}
-32c_5^2c_1(10c_1^2-c_5^2)\xi^2\Phi^2-2c_1^3c_5^2\Phi^2\big]\Phi_{\xi\xi}\\-224c_1^3c_5^2
\xi^3\Phi^2\Phi_{\xi\xi\xi}-32c_1^3c_5^2\xi^4\Phi^2\Phi_{\xi\xi\xi\xi}
-4c_5^3(-2c_5+c_1)(c_1+2c_5)\Phi^3=0
\end{multline}

Eq. \eqref{redphi} is a differential equation solely for $\Phi$ with variant coefficients. One can see that whence $\Phi$ is solved out from Eq. \eqref{redphi}, then $U,\, G$ and $H$ can be solved out directly from Eq. \eqref{reduphi}, \eqref{redgphi} and \eqref{redhphi}, respectively.
The explicit solutions for Boussinesq equation \eqref{bouss} is immediately obtained by substituting $U,H,G$ and $\Phi$ into Eq. \eqref{redusol05}.

\noindent\textbf{Case 2} $c_i\neq0$ (i=1,\,2,\,4) and $c_3=c_5=c_6=0$.

Similarly to case 1, solving the characteristic equation with the assumed condition, we get the following symmetry reduction solutions for the enlarged Boussinesq system
\begin{eqnarray}\label{redusol001}
\phi&=&\frac{c_1}{c_4\ln(\Delta)+c_1\Phi},\\
\label{redusol002}g&=&\frac{G}{\Delta(c_4\ln(\Delta)+c_1\Phi)^2},\\
\label{redusol003}h&=&\frac{2G^2}{c_1\Delta^2(c_4\ln(\Delta)+c_1\Phi)^3}+\frac{H}{\Delta^2(c_4\ln(\Delta)+c_1\Phi)^2},\\
\label{redusol004}m&=&\frac{M}{\Delta^2(c_4\ln(\Delta)+c_1\Phi)^2},\\
\label{redusol005}u&=&-\frac{2G^2}{c_1^2\Delta^2(c_4\ln(\Delta)+c_1\Phi)^2}-\frac{2H}{c_1\Delta^2(c_4\ln(\Delta)+c_1\Phi)}
+\frac{U}{\Delta^2},
\end{eqnarray}
with the group invariant variable $\xi$ unchanged as in case 1. Seeking the symmetry reduction equations for symmetry group invariant functions in this case, we substitute Eqs. \eqref{redusol001}, \eqref{redusol002}, \eqref{redusol003} and \eqref{redusol004} into eqs. \eqref{gh1}, \eqref{gh2}, \eqref{mphi} and find
\begin{equation}\label{2mphi}
-c_1^2\Phi_{\xi}-M=0,
\end{equation}
\begin{equation}\label{2gphi}
-c_4c_1+2c_1^2\xi\Phi_{\xi}-G=0,
\end{equation}
\begin{equation}\label{2gh}
2\xi G_{\xi}+G+H=0.
\end{equation}
Substituting Eq. \eqref{redusol001} into Eq. \eqref{schwarzform}, we have
\begin{multline}\label{2redphi}
(24c_1^3\xi^3+6c_1^3\xi)\Phi_{\xi}^3-(36c_1^2c_4\xi^2+64c_1^3\xi^6\Phi_{\xi\xi\xi\xi}
+c_4c_1^2+192c_1^3\xi^5\Phi_{\xi\xi\xi})\Phi_{\xi}^2+\big[288c_1^3\xi^5\Phi_{\xi\xi}^2
\\+(32c_1^2c_4\xi^3+256c_1^3\xi^6\Phi_{\xi\xi\xi}+4c_1^2c_4\xi)\Phi_{\xi\xi}+64c_1^2c_4
\xi^5\Phi_{\xi\xi\xi\xi}+320c_1^2c_4\xi^4\Phi_{\xi\xi\xi}-38c_1c_4^2\xi\big]\Phi_{\xi}\\-192c_1^3
\xi^6\Phi_{\xi\xi}^3-432c_1^2c_4\xi^4\Phi_{\xi\xi}^2-(128c_1^2c_4\xi^5\Phi_{\xi\xi\xi}
+160c_1c_4^2\xi^2+c_4^2c_1)\Phi_{\xi\xi}\\-16c_1c_4^2\xi^4\Phi_{\xi\xi\xi\xi}-112c_1c_4^2
\xi^3\Phi_{\xi\xi\xi}+c_4^3=0
\end{multline}
Now, substituting Eqs. \eqref{redusol001} and \eqref{redusol005} into Eq. \eqref{u2} with $u_2$ replaced by $u$ and vanishing $G,H,M$ by Eqs. \eqref{2mphi}, \eqref{2gphi}, \eqref{2gh}, we get
\begin{multline}\label{2uphi}
\big[12c_1^2(2U+7)\xi^2+c_1^2\big]\Phi_{\xi}^2+\big[144c_1^2\xi^3\Phi_{\xi\xi}+64c_1^2\xi^4\Phi_{\xi\xi\xi}
-4c_4c_1\xi(19+6U)\big]\Phi_{\xi}\\-48c_1^2\xi^4\Phi_{\xi\xi}^2-120c_4c_1\xi^2\Phi_{\xi\xi}
-32c_4c_1\xi^3\Phi_{\xi\xi\xi}+c_4^2(6U+5)=0
\end{multline}

From the symmetry reduction equations \eqref{2mphi}, \eqref{2gphi}, \eqref{2gh}, \eqref{2redphi} and \eqref{2uphi}, one can easily get the exact solutions of Boussinesq equation by substituting $U,\,G,\,H$ and $\Phi$ into Eq. \eqref{redusol005}.

It is seen that the symmetry reduction equation \eqref{2redphi} of $U$ with variant coefficient is still hard to solve. In order to give some explicit solutions of Boussinesq equation, we make some further constraints. For instance, we set $c_4=0$ and thus have
\begin{equation}
\Phi = \frac{16\sqrt{3}\Gamma(\frac{3}{4})^2BesselK(\frac{1}{4},\frac{\sqrt{3}}{8\xi})}{3\pi^2BesselI(\frac{1}{4},\frac{\sqrt{3}}{8\xi})}
\end{equation}
with integral constants taken as 0, then we finally get the special solution
\begin{equation}
u=\frac{c_1\big[-9BesselK(\frac{3}{4},\frac{\sqrt{3}}{8\xi})^2+5BesselK(\frac{1}{4},\frac{\sqrt{3}}{8\xi})^2\big]}{24BesselK(\frac{1}{4},\frac{\sqrt{3}}{8\xi})^2(c_1t+c_2)\xi}
\end{equation}

\noindent\textbf{Case 3} $c_i\neq0$ (i=2,\,3,\,4) and $c_1=c_5=c_6=0$.

Execute the similar procedure as in cases 1 and 2 under this condition, we can find the third group invariant solutions
\begin{eqnarray}\label{redusol}
\phi&=&\frac{2c_2}{2c_2\Phi+c_4(t-\xi)},\\
g&=&\frac{c_2^2G}{c_3^2(2c_2\Phi+c_4(t-\xi))^2},\\
h&=&\frac{c_2^2H}{c_3^2(2c_2\Phi+c_4(t-\xi))^2}+\frac{c_2^3G^2}{c_3^4(2c_2\Phi+c_4(t-\xi))^3},\\
m&=&\frac{c_2^2M}{c_3^2(2c_2\Phi+c_4(t-\xi))^2},\\
u&=&-\frac{c_2^2H}{c_3^2(2c_2\Phi+c_4(t-\xi))}-\frac{c_2^2G^2}{2c_3^4(2c_2\Phi+c_4(t-\xi))^2}+U,
\end{eqnarray}
with group invariant variable $\xi = -\frac{c_2x-c_3t}{c_3}$. The related symmetry reduction equations are
\begin{equation}
-4c_3^2\Phi_{\xi}-M=0
\end{equation}
\begin{equation}
4c_2^2\Phi_{\xi\xi}+H=0
\end{equation}
\begin{equation}
-2c_3c_4+4c_2c_3\Phi_{\xi}-G=0
\end{equation}
\begin{equation}
2c_3^2(6c_2^2U+c_3^2)\Phi_{\xi}^2+(8c_2^4\Phi_{\xi\xi\xi}-12c_3^2c_4c_2U)\Phi_{\xi}-6c_2^4
\Phi_{\xi\xi}^2-4c_4c_2^3\Phi_{\xi\xi\xi}+3c_3^2c_4^2U=0
\end{equation}
\begin{multline}
-128c_2^6\Phi_{\xi}^2\Phi_{\xi\xi\xi\xi}+\big[(128c_3^4c_4c_2+512c_2^6\Phi_{\xi\xi\xi}
)\Phi_{\xi\xi}+128c_4c_2^5\Phi_{\xi\xi\xi\xi}\big]\Phi_{\xi}-384c_2^6\Phi_{\xi\xi}^3
\\-(32c_3^4c_4^2+256c_4c_2^5\Phi_{\xi\xi\xi})\Phi_{\xi\xi}-32c_4^2c_2^4\Phi_{\xi\xi\xi\xi}=0
\end{multline}

Again, if we set $c_4=0$, then we have
\begin{equation}
\Phi=-\frac{2e_2}{e_1}\tanh\big(\frac{\xi+e_{3}}{2e_1}\big)
\end{equation}
with $e_1,\,e_2$ and $e_3$ being integral constants,
and
\begin{equation}
u = -\frac{(c_2^4+c_3^4e_1^2)\cosh(\frac{\xi+e_{3}}{2e_1}\big)^2+2c_2^4-c_3^4e_1^2}
{6c_3^2c_2^2e_1^2(\cosh\big(\frac{\xi+e_{3}}{2e_1}\big)^2-1)}
\end{equation}
as a special solution for Boussinesq equation.

\section{Conclusion and discussion}
In summary, Boussinesq equation is  investigated from the perspective of residual symmetry which is related with the truncated  Painlev\'{e} expansion. In order to apply residual symmetry to construct exact solutions of Boussinesq equation, we introduced multiple new variables such that the residual symmetry is localized in the new system. On the basis of the localized symmetry group, we have obtained B\"{a}cklund transformation for the enlarged Boussinesq system by using Lie's first theorem. After obtaining the general form of localized residual symmetry in the enlarged system, we considered the symmetry reduction in three subcases without  loss of generality. In these three cases under consideration, we obtained new symmetry reduction solutions for the enlarged Boussinesq system as well as the corresponding symmetry reduction equations for symmetry group invariant variables. Then the exact solutions for Boussinesq equation \eqref{bouss} can be constructed by solving these symmetry reduction equations. We also give some special explicit solutions under certain constraints.

In this paper, we have shown that nonlocal symmetry can be used to construct new exact solutions for nonlinear equations, so it is meaningful to get as much as possible nonlocal symmetries. Besides Painlev\'{e} analysis to obtain nonlocal residual symmetry, there exist other various ways to obtain nonlocal symmetries, such as those obtained from B\"{a}cklund transformation, the bilinear forms and negative hierarchies, the nonlinearizations \cite{cao,cheng} and self-consistent sources \cite{zeng} etc. Applying the localization procedure to these various nonlocal symmetries to constructed new exact solutions would deserve further study. Another critical question is what kind of nonlocal symmetries can be localized in a closed system, that also needs deep investigation to establish universal criterion for localizing procedure.

\begin{acknowledgments}
The authors are in debt to thank Prof. S.Y. Lou for his valuable comments and suggestions. This work was supported by the National Natural Science Foundation of China under Grant Nos. 10875078 and 11305106, the Natural Science Foundation of Zhejiang Province of China Grant Nos. Y7080455 and LQ13A050001.
\end{acknowledgments}

\end{document}